\documentclass[showpacs,twocolumn,floatfix,amsmath,amssymb,superscriptaddress,prb,nopacs]{revtex4-1}

\pdfoutput=1

\usepackage{amssymb}
\usepackage{epsfig}
\usepackage{graphicx}
\usepackage{color}

\newcommand{\sroa}{Sr$_{2}$RuO$_{4}$}
\newcommand{\sro}{SrRuO$_{3}$}

\newcommand{\bea}{\begin{eqnarray}}
\newcommand{\eea}{\end{eqnarray}}
\newcommand{\bt}{\textbf}

\begin{document}
\title{Inverse Proximity Effects at Spin-Triplet Superconductor-Ferromagnet Interface}
%
%Magnetic and superconducting proximity in SrRuO$_3$--Sr$_{2}$RuO$_{4}$ heterostructure}
%
\author{O. Maistrenko$^{*}$}
\affiliation{Max Planck Institute for Solid State Research,
D-70569 Stuttgart, Germany}
\author{C. Autieri$^{*}$}
\affiliation{International Research Centre MagTop, Institute of Physics, Polish Academy of Sciences, Aleja Lotnik\'ow 32/46, PL-02668 Warsaw, Poland}
\affiliation{SPIN-CNR, c/o Universit\`a di Salerno, IT-84084 Fisciano (SA), Italy}
\author{G. Livanas}
\affiliation{SPIN-CNR, c/o Universit\`a di Salerno, IT-84084 Fisciano (SA), Italy}
\affiliation{Department of Physics, National Technical University of Athens, GR-15780 Athens, Greece},
\author{P. Gentile}
\affiliation{SPIN-CNR, c/o Universit\`a di Salerno, IT-84084 Fisciano (SA), Italy}
\affiliation{Dipartimento di Fisica ``E. R. Caianiello", Universit\`a di Salerno, IT-84084 Fisciano (SA), Italy}
\author{A. Romano}
\affiliation{Dipartimento di Fisica ``E. R. Caianiello", Universit\`a di Salerno, IT-84084 Fisciano (SA), Italy}
\affiliation{SPIN-CNR, c/o Universit\`a di Salerno, IT-84084 Fisciano (SA), Italy}
\author{C. Noce}
\affiliation{Dipartimento di Fisica ``E. R. Caianiello", Universit\`a di Salerno, IT-84084 Fisciano (SA), Italy}
\affiliation{SPIN-CNR, c/o Universit\`a di Salerno, IT-84084 Fisciano (SA), Italy}
\author{D. Manske}
\affiliation{Max Planck Institute for Solid State Research,
D-70569 Stuttgart, Germany}
\author{M. Cuoco}
\affiliation{SPIN-CNR, c/o Universit\`a di Salerno, IT-84084 Fisciano (SA), Italy}
\affiliation{Dipartimento di Fisica ``E. R. Caianiello", Universit\`a di Salerno, IT-84084 Fisciano (SA), Italy}
\begin{abstract}
We investigate inverse proximity effects in a spin-triplet superconductor (TSC) interfaced with a ferromagnet (FM), assuming different types of magnetic profiles and chiral or helical pairings. The region of the coexistence of spin-triplet superconductivity and magnetism is significantly influenced by the orientation and spatial extension of the magnetization with respect to the spin configuration of the Cooper pairs, resulting into clearcut anisotropy signatures. A characteristic mark of the inverse proximity effect arises in the induced spin-polarization at the TSC interface. This is unexpectedly stronger when the magnetic proximity is weaker, thus unveiling immediate detection signatures for spin-triplet pairs. We show that an anomalous magnetic proximity can occur at the interface between the itinerant ferromagnet, SrRuO$_3$, and the unconventional superconductor Sr$_2$RuO$_4$. Such scenario indicates the potential to design characteristic inverse proximity effects in experimentally available SrRuO$_3$-Sr$_2$RuO$_4$ heterostructures and to assess the occurrence of spin-triplet pairs in the highly debated superconducting phase of Sr$_2$RuO$_4$.  
\end{abstract}
\maketitle

{\it Introduction.} 
Spin-triplet superconductors can be odd-parity in momentum, e.g. $p$-wave~\cite{Sigrist,TanakaPWave,ReadGreen,Ivanov,Kitaev01,VolovikBook,Maeno2012,SatoAndo}, or even parity for multi-orbital configurations ~\cite{Spalek2001,Puetter2012,Han2004,Fukaya2018,Suh2020} Apart from the variety of orbital pairings, the spin-triplet part of the superconducting (SC) order parameter (OP), typically encoded in a ${\bf d}$-vector, has a substantial imprint on the superconducting properties. In fact, the spin degree of freedom in superconductors naturally enriches the physical scenario resulting into anomalous response to Zeeman or ferromagnetic fields~\cite{Murakami,Dumi1,Dumi2,Wright,Mercaldo2016,Mercaldo2017,Mercaldo2018}, spin-sensitive Josephson transport~\cite{Sengupta,Yakovenko,Yakovenko1,Cuoco1,Cuoco2,BrydonJunction,Gentile,Brydon1,Brydon2}, magnetic topological reconstructions~{\cite{Mercaldo2016,Mercaldo2017,Mercaldo2018}, and, on a general ground, may lead to energy efficient superconducting spintronics~\cite{Linder2015}. Notably, intrinsic or engineered spin-triplet superconductors can host Majorana bound states and thus are particularly impactful for topological quantum computing~\cite{Ivanov,Kitaev03,Nayak,Alicea}.

While more rare in nature than the canonical spin-singlet one, several experimental observations have led to evidences for spin-triplet superconductivity in a large variety of materials~\cite{Stewart,Saxena,Kyogaku,Jiao,Ran,Bauer,Nishiyama,Lebed,Lee,
Shinagawa,Bao15,Cuono19,Noce2020}. A notable case is Sr$_2$RuO$_4$ whose superconducting nature is under intense debate because recent measurements~\cite{Pustogow2019,Ishida2020,Petsch2020} pose serious constraints on the long thought spin-triplet chiral $p$-wave superconductivity \cite{Mackenzie2017}.
%proposed to characterize that system.
Such case underlines the remarkable challenges that are typically encountered in assessing the spin-triplet pairing in unconventional superconductors. 

In this work we take an alternative exploratory path to access the spin-triplet nature of the superconducting pairing by focusing on its spin degree of freedom and investigating the inverse proximity effects that can occur at the spin-triplet superconductor (TSC)-ferromagnet (FM) interface, assuming that spin polarization may leak into the TSC. Inverse proximity effects have been studied in heterostructures involving spin-singlet superconductors ~\cite{Tokuyasu1988,Bergeret2004,Tollis2015,
Bergeret2005} with unexpected consequences when nodal excitations occur~\cite{Dibernardo2019}.
Instead, while it is well consolidated that the physics of TSC-FM heterostructures is richer than their singlet counterparts because the orientation of the FM moment relative to the TSC {\bt{d}}-vector introduces extra channels of coupling, to the best of our knowledge, the inverse proximity effects both for the magnetic and superconducting components in TSC-FM have not yet been fully addressed.
There are various fundamental physical aspects to be accounted for once the spin polarization of the FM penetrates into the TSC regarding the reconstruction of the SC-OP and the modification of the magnetic proximity (MP) due to the presence of the spin-triplet pairs. The analysis is performed by considering two prototypical spin-triplet pairings with uniaxial and planar \bt{d}-vector and various characteristic spatial profiles of the magnetization in the TSC with inequivalent character of the MP (Fig. \ref{f1}). We find that the TSC order parameter in the region of the magnetic-superconducting coexistence is significantly sensitive to the orientation of the magnetization, thus unveiling clearcut anisotropy signatures. Similarly, the behavior of the induced spin-polarization in the TSC depends on the character of the \bt{d}-vector and on the inverse MP pattern. Then, motivated by the timely debate on the nature of the superconducting phase in Sr$_2$RuO$_4$ \cite{Pustogow2019,Petsch2020} and on the anomalous magneto-transport properties of SrRuO$_3$-Sr$_2$RuO$_4$ heterostructure \cite{Anwar2016,Anwar2019a,Anwar2019b}, we evaluate the MP of the Sr$_2$RuO$_4$ when interfaced with SrRuO$_3$. By ab-initio methods we find that the SrRuO$_3$ induces a large magnetic moment in the Sr$_2$RuO$_4$ interface layers with characteristic profiles that depend on the strength of the Coulomb interaction. This magnetic scenario supports the general investigation of the inverse proximity effects in the presence of different types of magnetic leakage into the TSC. 
The recent developments in the fabrication of Sr$_2$RuO$_4$ thin films~\cite{Krock2010,Marshall2017,Uchida2017,Palomares2020} and hetero-structures allow to experimentally investigate the properties of the inverse proximity effect, which we discuss in our work. This can provide relevant experimental paths to address the debate about the nature of the superconductivity in Sr$_2$RuO$_4$.

\begin{figure}[bt]
\includegraphics[width=0.44\textwidth]{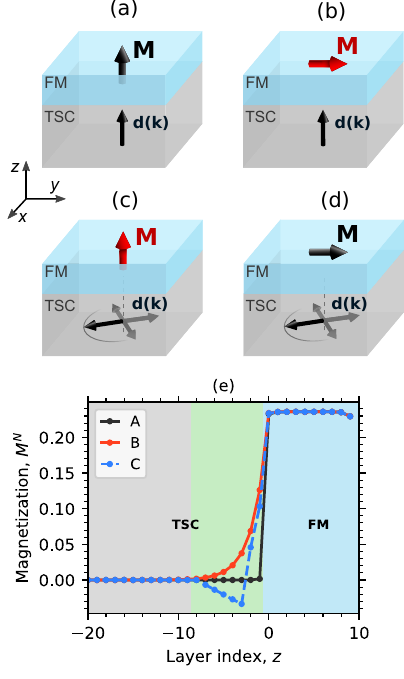} 
%\hspace{0.4cm}
\protect\caption{Illustration of the magnetic and superconducting configurations for the TSC-FM heterostructure with chiral (a),(b) and helical (c),(d) order parameters, respectively. The magnetization is out-of-plane (a),(c) or in-plane (b),(d) oriented. (e) the examined magnetic profiles can have (A) a step-like magnetization at the interface with substantial absence of magnetic proximity, (B) a monotonous decaying of the spin polarization within the TSC on the scale of the superconducting coherence length, or (C) a sign changing penetration in the TSC, respectively. The blue region stands for the ferromagnetic layers. In the gray area the spin-polarization induced by the FM into the TSC has vanishing amplitude.}
\label{f1}
\end{figure}

\begin{figure*}[bt]
\includegraphics[width=0.94\textwidth]{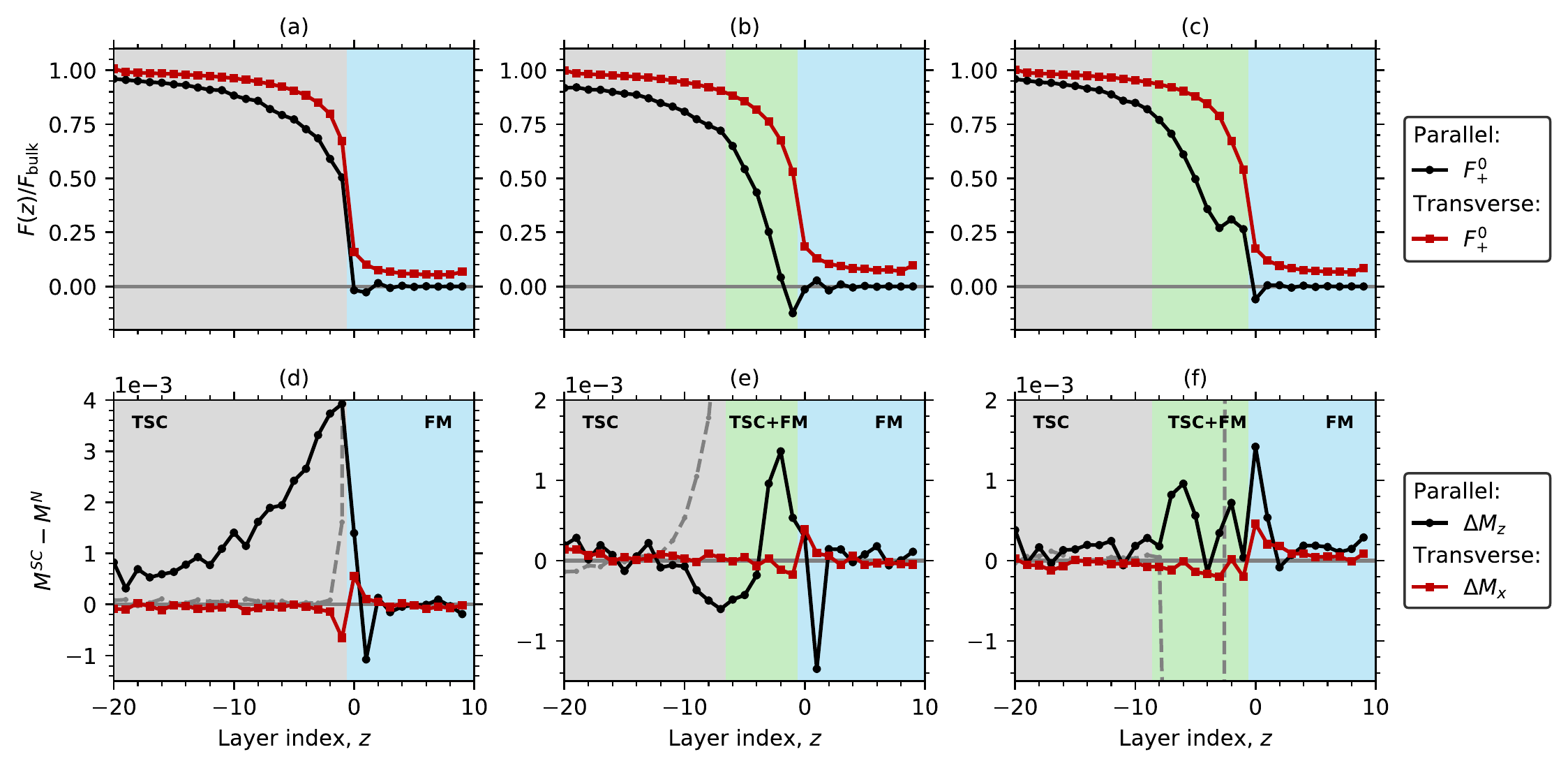}  
%\hspace{0.4cm}
\protect\caption{Evolution of the pairing amplitude for chiral TSC heterostructure near the TSC-FM interface assuming different magnetization profiles as reported in Fig. \ref{f1}(e). (a),(b) and (c) correspond to step-like, monotonous decay, and sign-changing spatial dependence of the magnetization, respectively. The black circles (red squares) refers to a configuration with magnetization oriented along the out-of-plane $z$ -direction (in-plane $x$-direction), respectively. The $z$-($x$-) spin polarization is collinear (transverse) to the ${\mathbf{d}}$-vector. In the panels (d),(e),(f) we show the spatial dependence of the difference, $\Delta M$, between the magnetization in the superconducting ($M^{SC}$) and normal state ($M^N$) for the parallel (circles) and transverse (squares) configurations, respectively. 
%Here, (a),(b) and (c) correspond to step-like, monotonous decay, and sign-changing spatial dependence of the magnetization, respectively. 
The dashed line in (d),(e),(f) indicates the amplitude of the magnetization in the normal metal state. The green area indicates the layers where the magnetization penetrates into the TSC and there is a coexistence of non-vanishing magnetization and superconductivity. The blue region stands for the ferromagnetic layers. In the gray area the spin-polarization induced by the FM is substantially zero.}
\label{f2}
\end{figure*}

%%%
{\it Model and methodology.} We consider an FM-TSC heterostructure with a layered geometry described by a single-band tight-binding model with a spin dependent nearest-neighbor attractive interaction that can stabilize spin-triplet pairing with chiral or helical $\bt{d}$-vectors (Fig. 1) ~\cite{Cuoco1,Cuoco2,Terrade2014,Terrade2013}. The total Hamiltonian of the system is then defined on a lattice with size
$L_x\times L_y \times L_z$ ($L_x=L_y=L$) assuming periodic boundary
conditions along $x$ and $y$. The simulation is performed for a
system having $L=100$ and $L_z=40$. Since the out-of-plane  superconducting coherence length is of the order of about six unit cells, the size of the system is adequate for tracking the inverse and direct proximity effects on the characteristic superconducting length scales.  We indicate each site by a
vector $\vec{i} \equiv(i,i_z)$, with $i$ labeling the $(xy)$
in-plane atomic positions and $i_z$ the different layers along the
$z$-direction. The FM region corresponds to layers with $i_z \geq 0$ and the TSC one to layers with $i_z<0$, respectively. The Hamiltonian can be expressed as

%%%%%
\begin{eqnarray}
&H& = \sum_{{\mathbf k},i_{z},\sigma} (\epsilon_{{\mathbf k} \sigma} -\mu)
c^{\dagger}_{{\mathbf k} \sigma}(i_z) c_{{\mathbf k} \sigma}(i_z) 
- \sum_{i_z}
{\mathbf{h}}(i_z) \cdot {\mathbf{M}}(i_z)+
\nonumber
\\&-& \sum_{<i j>, i_z<0 } V_{\sigma,\sigma'} n_{i \sigma}(i_z)
n_{j \sigma'}(i_z) \nonumber \\
&+& \sum_{\delta=\pm1} \sum_{{\mathbf k} \sigma} t_{\perp} (c^{\dagger}_{{\mathbf k}
\sigma}(i_z + \delta) c^{}_{{\mathbf k} \sigma}(i_z) + {\text{h.c.}}) %\nonumber \\
\label{ham}
\end{eqnarray}
%%%%
%
\noindent with $c_{{\bf k}\sigma}(i_z)$ being the annihilation operator of an electron
with planar momentum ${\bf k}$, spin $\sigma$ at the layer $i_z$ and $\langle i j\rangle $ denotes the in-plane nearest-neighbor
sites, $\mu$ is the chemical potential, $V_{\sigma,\sigma'}$ is the spin-dependent in-plane pairing strength, ${\mathbf{M}}(i_z) =
\sum_{i,s,s'}c^{\dagger}_{s}(i){\vec{\sigma}}_{s,s'}c_{s'}(i)$
is the spin density of the layer $i_z$, and $\epsilon_{k
\sigma}=-2 t [\rm{cos}(k_x)+\rm{cos}(k_y)]$ is the in-plane electronic
spectrum. ${\mathbf{h}}(i_z)$ is the layer dependent exchange field that sets the amplitude and orientation of the magnetization within the heterostructure (Fig. \ref{f1}). We assume a good electronic matching at the interface expressed by having $t_{\perp}=0.5 t$. The selection of the three profiles in Fig. \ref{f1} is general enough to include the phenomenology which is expected when interfacing a correlated paramagnetic metal with a ferromagnet. In fact, the breaking of the translational symmetry due to the interface allows to get a reconstruction of the MP with monotonous or oscillating spatial behavior. Such occurrence is also expected to be most likely to occur in paramagnetic electronic systems having a peak at finite momentum in the magnetic suceptibility. Furthermore, since the Fermi length is of the order of the unit cell dimension, the MP is taken as more significant in the layers nearby the interface.\\

For the present study the quartic term
of the model Hamiltonian in Eq. \ref{ham} is decoupled in the canonical
Hartree-Fock approximation within the pairing channel. We
introduce the layer dependent superconducting pairing amplitude for each spin-triplet channel to self-consistently determine the corresponding order parameters with chiral and helical symmetry as:
\begin{eqnarray}
F^{0}_{\theta}&=&\frac{1}{L^2} \sum_{\sigma \neq \sigma',{\mathbf{k}}} (\sin k_x -i \theta \sin k_y) \langle c_{-{\mathbf{k}}\sigma} c_{{\mathbf{k}}\sigma'} \rangle \nonumber \\ 
F^{\sigma}_{\theta}&=&\frac{1}{L^2} \sum_{\mathbf{k}} (\sin k_x -i \theta \sin k_y) \langle c_{-{\mathbf{k}}\sigma} c_{{\mathbf{k}}\sigma} \rangle \nonumber 
\end{eqnarray} 
\noindent where $\theta=\pm$ is employed to set the winding of the spin dependent order parameter, while $\langle ... \rangle$ is the average over the ground state. Depending on the ampltitudes of $V_{\sigma,\sigma'}$ the chiral or helical spin-triplet superconductivity can be stabilized \cite{Terrade2013,Cuoco1,Gentile}. 

\begin{figure*}[bt]
\includegraphics[width=0.94\textwidth]{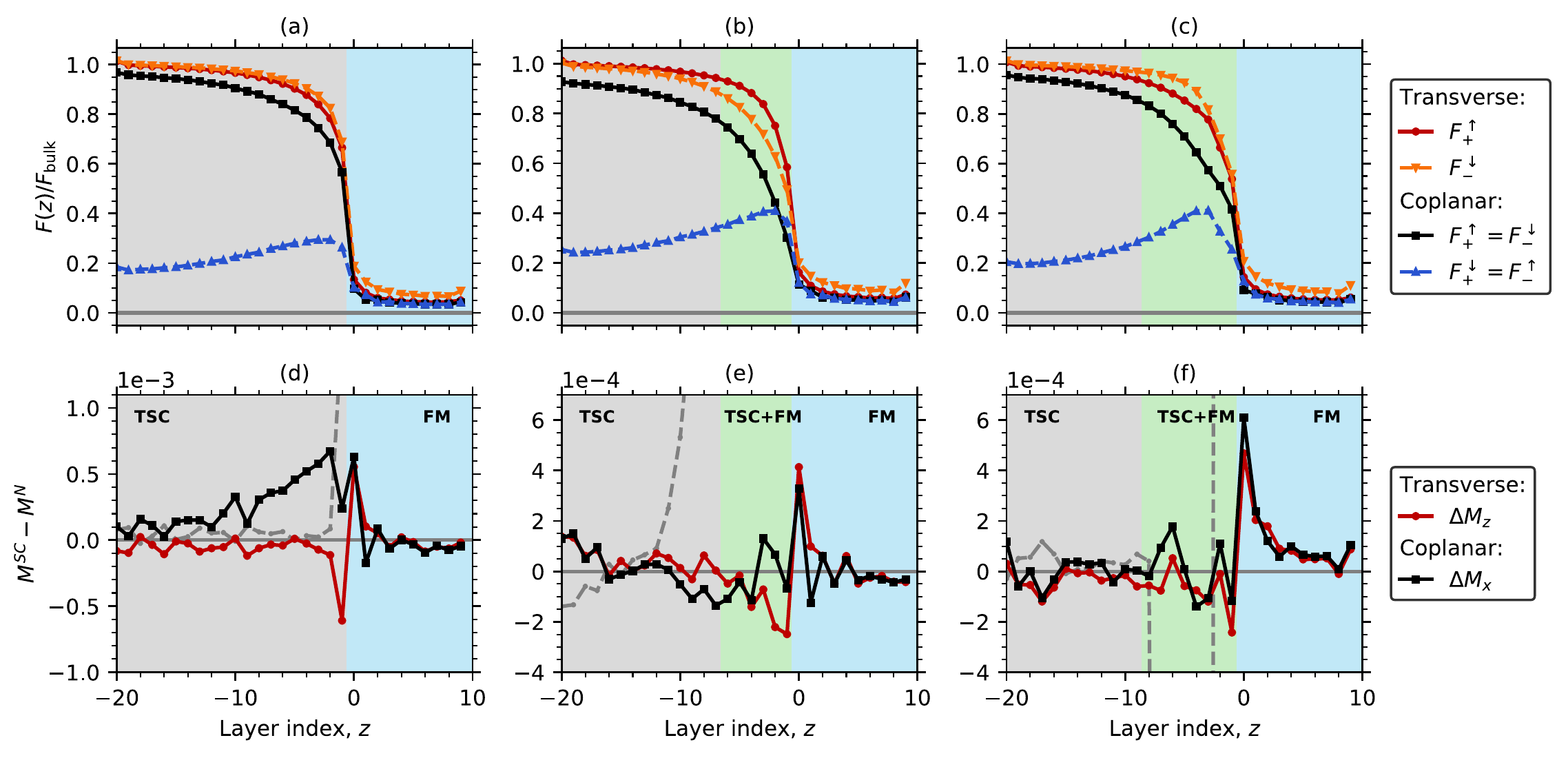}  
%\hspace{0.4cm}
\protect\caption{Evolution of the pairing amplitude for the helical-TSC heterostructure near the TSC-FM interface, assuming different magnetization profiles in the TSC, as reported in Fig. \ref{f1}(e). (a),(b) and (c) corresponds to the step-like, monotonous decay, and sign-changing spatial dependence of the magnetization, respectively. Circles (squares) stand for a magnetization which is oriented along the out-of-plane $z$-direction (in-plane $x$-direction), respectively. The $z$-($x$-) oriented spin polarization is transverse (coplanar) to the ${\mathbf{d}}$-vector. In the panels (d),(e),(f) we report the spatial dependence of the difference, $\Delta M$, between the magnetization in the superconducting ($M^{SC}$) and in the normal ($M^{N}$) state with coplanar and transverse configurations. The dashed gray line in (d),(e),(f) indicates the amplitude of the magnetization in the normal metal for the different types of examined magnetic proximity.}
\label{f3}
\end{figure*}

{\it Inverse superconducting and magnetic proximity effects.}  
We start by discussing the superconducting proximity effects for the chiral $d_z$-TSC (Fig. \ref{f2}), assuming three different magnetization leakage into the TSC (Fig. 1 (e)). The chiral amplitude $F^{0}_{+}$ depends on the orientation of the FM spin-polarization (Fig. \ref{f2}(a)), exhibiting a suppression of its value when the magnetization is parallel ($M_z$) to the ${\mathbf d}$-vector compared to the transverse configuration ($M_x$), as schematically shown in Figs. \ref{f1} (a,b).

Here, the use of different MP profiles allows us to assess how the leakage of spin polarization in the TSC affects the SC-OP and the resulting spin polarizability of the TSC at the interface. 

Firstly, we observe that in the magnetized region of the TSC, the amplitude of the chiral order parameters decreases for the $M_z$ orientation, while it is substantially unaffected for the transverse $M_x$ case. This is consistent with the spin orientation of the Cooper pairs.    
The behavior with transverse spin-polarization $M_x$-FM (Fig. \ref{f2}(b),(c)) unveils a weak enhancement when the inverse MP increases in length penetration. Such outcome substantially indicates that the spin-polarization entering into the TSC yields more spin-triplet pairs for the proximity into the FM.  
For the $M_z$-state we notice that the expected damped oscillation of the pairing amplitude does not exhibit variation as a function of the MP. 

Let us now focus on the behavior of the induced spin polarization due to the leakage or lack of magnetization within the TSC region of the heterostructure. From the behavior of the SC order parameter it is immediate to observe that the effects in the TSC are tied to the orientation of the magnetization with respect to the ${\mathbf{d}}$-vector. Then, in order to relate the behavior of the induced magnetization to the character of the MP,
the strategy is to compute the magnetization difference for each layer of the superconducting state with respect to that of the normal metal configuration, i.e. that one corresponding to a vanishing pairing interaction $V$. This physical quantity gives indication of the response of the superconductor close to the interface not only for the presence of a non-vanishing exchange field in the region where the order parameter is typically reduced in amplitude, but also to the gradient of the spin-polarization which is particularly enhanced at the TSC-FM boundary. In Figs. \ref{f2}(d),(e),(f) we follow the spatial evolution of the induced magnetization for the various MP profiles. As a general trend, we observe that the magnetization change between the superconducting and the normal states is substantially negligible for the $M_x$ orientation, i.e. transverse to the ${\mathbf{d}}$-vector with the exception of the layers very close to the interface. There, the induced magnetization  typically changes sign irrespective of the MP profile, thus indicating that the TSC tends to screen the MP by spatially modulating the spin-polarization. For the transverse magnetization, it is interesting to link the enhancement of the SC proximity effect (Fig. \ref{f2}(c)) with that of the spin polarization in Fig. \ref{f2}(f). The result indicates that an increase of the magnetization can be observed at distance of several layers from the interface due to the  penetration of the spin-triplet Cooper pairs with spin polarization collinear to $M$.
On the other hand, the response of the TSC to the $M_z$ leakage, i.e. parallel to the ${\mathbf{d}}$-vector, is typically larger in amplitude and significantly sensitive to the character of the MP. We find that for the case of a step-like magnetization at the interface the response of the TSC is paramagnetic (Fig. \ref{f2}(d), i.e. the spin polarization is enhanced with respect to the normal state), with a penetration in the superconductor over a spatial range of the order of the out-of-plane coherence length. The impact of the MP is significantly different when considering the case of a longer range penetration of the magnetization in the TSC with monotonous decay (Fig. \ref{f2}(e)). 
In this regime, the amplitude of the induced spin-polarization spatially oscillates around the one in the normal state configuration. Thus, the TSC yields a vanishing net spin polarization in average. When considering the sign-changing MP (Fig. 2(f)), we find that the magnetic reconstruction due to the TSC leads to an overall paramagnetic response with an enhancement of the spin polarization distribution in all layers close to the interface. We argue that the variation of the induced magnetic effects is mainly due to the gradient of the magnetization close to the interface because the amplitude is comparable for the two configurations.  
In fact, the character of the observed induced spin-polarization nearby the TSC-FM interface can be grasped by analyzing a simplified effective model with only two superconducting layers in the presence of an inhomogeneous exchange field. The details of the outcome of such effective model are reported in the Appendix. The changeover from paramagnetic to diamagnetic of the TSC response as a function of the layer position can be ascribed to the gradient of the exchange field rather than to the effective strength of the magnetization with respect to the SC gap (see Appendix).

At this stage it is worth asking whether and how the magnetic reconstruction at the TSC-FM gets modified if one is considering a TSC with helical pairing, i.e. $F^0=0$ and $F^{\uparrow}_{+}=F^{\downarrow}_{-}$ thus marked by a two-components ${\mathbf{d}}$-vector lying in the $xy$ plane, as shown in Fig. \ref{f1} (c),(d). For the helical TSC the superconducting proximity does not exhibit an oscillating behavior because the ${\mathbf{d}}$-vector lies in a plane and thus one cannot find a magnetization direction that is fully collinear to it across the whole Brillouin zone. The behavior of the TSC pairing amplitude is not much affected by the change of the MP (Fig. \ref{f3}(a),(b),(c)). Moreover, for the coplanar configuration the presence of the magnetic interface induces a pairing amplitude in each spin-channel with opposite winding compared to that of the SC order parameter, and this is irrespective of the character of the MP. Such component can yield a transverse contribution in the spin polarization with respect to the magnetization orientation. More interesting is the behavior of the spin polarization in the TSC. One finds that for the step-like magnetization profile, the response of the TSC is paramagnetic-like for the coplanar configuration while it is diamagnetic-like for the transverse $M_z$ state (see Fig. 3(d)). The effect typically extends over a distance of the order of the coherence length. The amplitude of the induced spin polarization is approximately isotropic. This aspect is a consequence of the multi-component ${\mathbf{d}}$-vector.  A similar isotropic behavior is also observed for the case of longer range penetration of the magnetization with a tendency to exhibit a sign change both in the TSC and in the FM side of the heterostructure (Fig. \ref{f3}(e),(f)).

{\it Magnetic proximity at the Sr$_2$RuO$_4$-SrRuO$_3$ interface.} Here, in order to have an experimentally relevant case of inverse magnetic effects, we consider the Sr$_2$RuO$_4$-SrRuO$_3$ heterostructure. SrRuO$_3$ is an itinerant ferromagnet, which over the past years has
been the subject of intense research \cite{Matsuno2016,Ohuchi2018,Li2020,Kan2020,Groenendijk2020}.   
By means of the DFT approach (see Appendix for details) we demonstrate that the magnetic behavior of the Sr$_2$RuO$_4$-SrRuO$_3$ is non-standard if compared to conventional metal-ferromagnet interface and can lead to magnetization profiles resembling those that have been employed in the present study. Starting from the result that the bulk Sr$_2$RuO$_4$ (see Appendix for details) is paramagnetic when the Coulomb repulsion $U_{214}$ is below about 0.7~eV, we find that for the interface layers proximized to SrRuO$_3$ the Coulomb threshold for the transition into the ferromagnetic state is significantly reduced (see Appendix). Hence, to evaluate how the ground state of the Sr$_2$RuO$_4$ is affected by the ferromagnetism in SrRuO$_3$ we investigate a superlattice made of five Ru-O layers in each side of the heterostructure (Fig. \ref{f4}(a)). 
We find that the ground state of the supercell is generally ferromagnetic in {\sro} with a magnetization that slightly decreases at the interface layers. {\sroa}~ can also lower its energy by developing a non-vanishing magnetization that is collinear to that of the {\sro} close to the interface and can change sign away from it depending on the strength of the Coulomb interaction (Fig. \ref{f4}(b)).
The interface layer has always the largest magnetization within {\sroa}.
We tested several different magnetic orderings for the SrRuO$_3$-Sr$_{2}$RuO$_{4}$ supercell, finding that the ferromagnetic coupling 
between SrRuO$_3$ and Sr$_{2}$RuO$_{4}$ turns out to be the most stable for the supercell.
For U$^{214}$ in the energy range [0.4-0.6]~eV, we have a sign-changing magnetization in the layers of Sr$_{2}$RuO$_{4}$
while for U$_{214}$=0.7 eV all layers have a spin polarization which is aligned to that of the \sro~and progressively decay in the Sr$_{2}$RuO$_{4}$ side of the heterostructure.
The local magnetic moment increases with U$_{214}$, as expected, while
the magnetic moment of SrRuO$_3$ stays substantially unaltered.
Although the computational time limits the size of the supercell which can be simulated, the resulting behavior provides clear trends of the way magnetization gets reconstructed nearby the \sro-\sroa~ interface.

\begin{figure}[bt]
\includegraphics[width=0.475\textwidth]{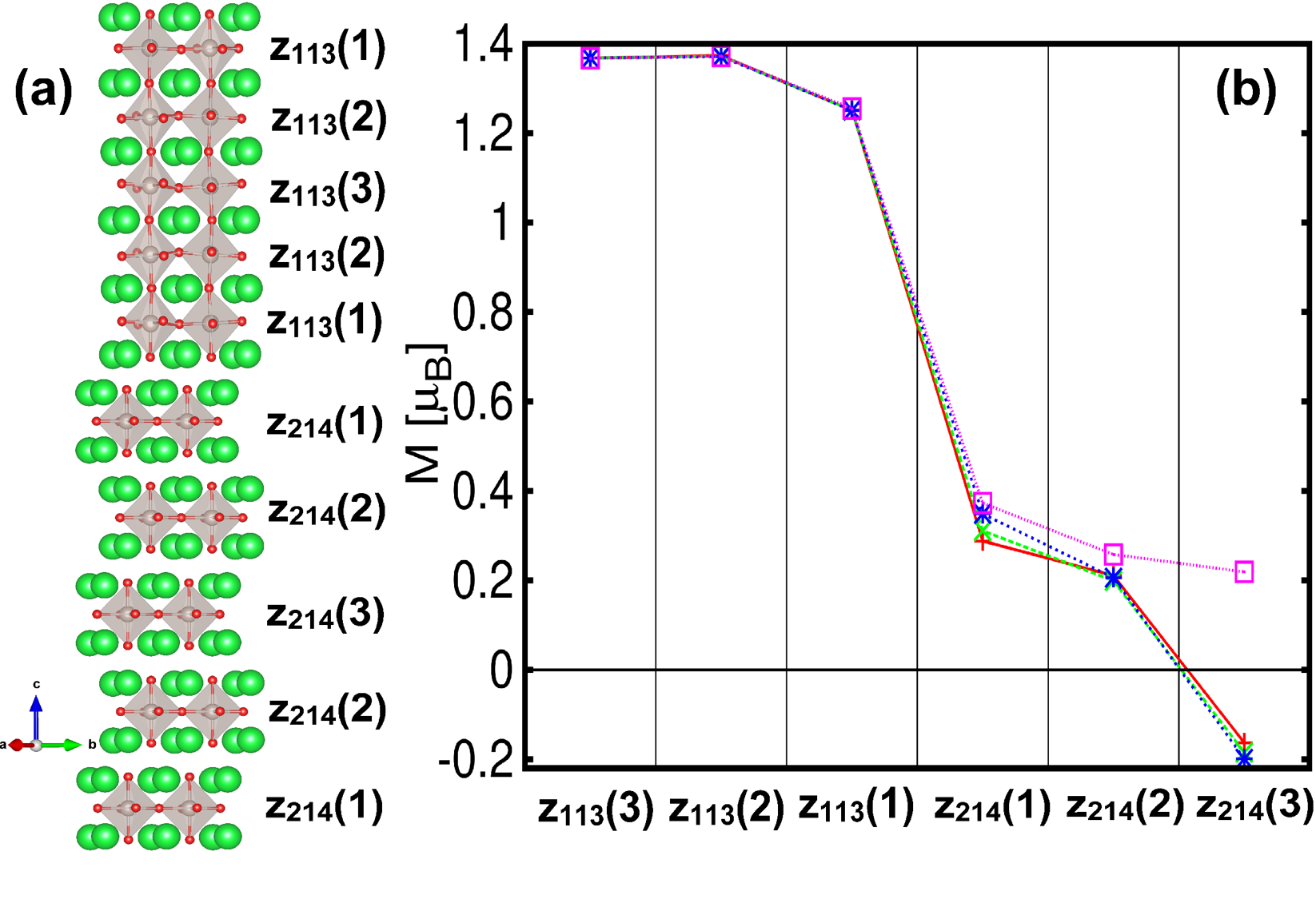}  
%\hspace{0.4cm}
\protect\caption{(a) Schematic illustration of the {\sro}-{\sroa} superlattice with five Ru-O layers having cubic (\sro) and tetragonal structure (\sroa), respectively. (b) evolution of the magnetization across the interface for various values of the Coulomb interaction $U_{214}$ in \sroa. Red, green, blue and magenta points are
for $U_{214}$=0.4,0.5, 0.6 and 0.7~eV, respectively. z$_{113}(n)$ and z$_{214}(n)$ are the n-th layer from the interface for the corresponding {\sroa} and {\sro} sides of the superlattice.}
\label{f4}
\end{figure}

{\it Conclusions.} We have demonstrated that inverse proximity effects in TSC-FM heterostructure exhibit clearcut signatures for both magnetic and superconducting interface reconstruction with respect to the nature of the pairing and the character of the magnetization leakage in the TSC. The uniaxial ${\bf d}$-vector for the chiral TSC shows a magnetic response which is strongly anisotropic, with a tendency to a significant spatial modulation in amplitude and sign for the case of a long-range MP (Fig. \ref{f2}). The helical TSC with planar ${\bf d}$-vector tends instead to yield a more isotropic spin-polarization, except for the step-like magnetic profile, which is smaller (larger) in amplitude than that of the chiral state in the TSC (FM) side of the junction (Fig. \ref{f3}). These features can be relevant to account for the observed anisotropy of the transport properties in Sr$_2$RuO$_4$-SrRuO$_3$ \cite{Anwar2019a,Anwar2019b}. In such a system, by means of ab-initio approaches we demonstrate that there occurs a magnetic instability near the interface for the electronic states of Sr$_2$RuO$_4$. We argue that the tendency to undergo a changeover from paramagnetic to ferromagnetic in the Sr$_2$RuO$_4$ can set out the underlying scenario for the anomalous enhancement of the magnetic moments in Sr$_2$RuO$_4$-SrRuO$_3$ heterostructure \cite{Sugimoto2015}. Our results indicate that the presence of spin-triplet pairs can be accessed through magnetic measurements that are spatially resolved with respect to the TSC-FM interface and performed across the superconducting transition. By tracking the relation of the spatially resolved magnetization in the normal and the superconducting state one can get unique hallmarks of the $\bt{d}$-vector structure in the superconductor. 

\begin{acknowledgments}
$^{*}$These authors contributed equally to this work.  
C. A. is supported by the Foundation for Polish Science through the International Research
Agendas program co-financed by the European Union within the Smart Growth Operational Programme. 
We acknowledge the CINECA award under the
ISCRA initiatives IsC76 "MEPBI" and IsC81 "DISTANCE"
Grant, for the availability of high-performance computing
resources and support.
\end{acknowledgments}

\appendix

\section{Effective model for magnetic proximity in the spin-triplet superconductor}

In this Appendix, we compare the impact of the exchange field on both mono-layer and bilayer geometries with chiral and helical spin-triplet superconductivity in order to evaluate in a simplified configuration the local role of an inhomogeneous magnetic proximity close to the interface. 

\subsection{Superconducting spin polarization vs exchange field for monolayer}

Let us start by considering a mono-layer spin-triplet superconductor in the presence of a Zeeman term that mimics the induced magnetic proximity.
We investigate the behavior of the normal state magnetization $M^{\text{N}}$ compared with the spin-polarization $M^{\text{SC}}$ induced by the magnetic exchange in the superconducting state. This analysis allows to understand  how the induced spin-polarization evolves in the spin-triplet superconductor. 

In the case of chiral superconductivity, the $\bf{d}$-vector is uniaxial and pointing along the out-of-plane direction with respect to the electron motion. This means that Cooper pairs have $S_z=0$,
and the equal spin configurations are generically aligned in the $x-y$ plane.
%If we have $h_z$ field ($\parallel \vec{d}$) it can not polarize Cooper pairs,
%and at zero temperature $M^{SC}$ is zero (system shows strong diamagnetic behavior).
%Only for the field $h_z > \Delta_0$ the superconductivity is killed due to the Pauli limiting and the system will show normal state behavior. 
The outcome of the spin-polarization in the superconducting state is reported in Tables~\ref{table:single-layer-chiral},\ref{table:single-layer-helical}. We find that the induced spin moments are always aligned to those of the normal-state magnetization and, as expected, there is a tendency to have a vanishing spin polarization with the field parallel to the $\bt{d}$-vector upon reaching an amplitude that is comparable to the superconducting gap.
Instead, if we have a configuration of $\bt{h}$ perpendicular to the $\bt{d}$-vector than there is always a net spin polarization due to the Cooper pair spin moments being aligned to $\bt{h}$.
Hence, Cooper pairs can get spin polarized without pair breaking and the magnetization has a typical amplitude that is close to that of the normal state.

In the case of helical superconductor, the $\bt{d}$-vector has a planar orientation that changes in the momentum space. This means that there are Cooper pairs with both spin projections $S_x=0$ and $S_y=0$. This results into a superconducting order parameter with Cooper pairs that are spin polarized along the $z$-direction.
Starting with an exchange field that is transverse to the $\bt{d}$-vector, we have that the spin moment of the Cooper pairs is aligned to $\bt{h}$.
Hence, we obtain $M^{\text{SC}}=M^{\text{N}}$ as it was for the chiral spin-triplet pairing.
The case with $h_x$ is richer in terms of possible outcomes. 
For this configuration there are $\vec{k}$ points in the reciprocal space for which one gets $\bt{h}$ parallel to $\bt{d}$,
while for others $\bt{h}$ is noncollinear to $\bt{d}$.
Indeed, one has an angle between the field and the superconducting $\bt{d}$-vector  that can vary between $0$ and $\pi$.
Then, for $h_x < \Delta$ one can only spin polarize a portion of spin triplet pairs, and consequently $M^{\text{SC}} < M^{\text{N}}$.
Additionally, by allowing a configuration of the order parameter with $C_4$ rotational symmetry breaking, we obtain that the superconductor can gain energy by having more electron pairs polarized at the expenses of redistributing them to the Fermi surface, making the order parameter asymmetric, i.e. $d_x \neq d_y$.
In this case, we obtain an increase of the magnetization $M^{\text{SC}^{*}} > M^{\text{SC}}$. 
The resulting findings are summarized in Table~\ref{table:single-layer-helical}.

\begin{table}[ht]
%\caption{Chiral}
\begin{tabular}{ |l|c|c| }
      \hline
      & $\bt{h} \parallel \bt{d}$ & $\bt{h} \perp \bt{d}$ \\
      \hline
      $h / \Delta_0$        & $M^{\text{SC}}$                    & $M^{\text{SC}}$                \\
      \hline
      a) $h < \Delta$       & 0                           & $0.9 M^{\text{N}}$             \\
      b) $h \approx \Delta$ & 0                           & $M^{\text{N}}$                   \\
      c) $h > \Delta$       & $M^{\text{N}}$                     & $M^{\text{N}}$                   \\
      \hline
\end{tabular}
\caption{Summary of the induced magnetization $M^{\text{SC}}$ for a monolayer with chiral superconductivity. Results are for two directions of the magnetic exchange field in three different regimes of amplitude for $\bt{h}$.}
\label{table:single-layer-chiral}
\end{table}
    
%%%%%%%%%%%%%%%%%%%%   
%%%%%   Helical   %%
%%%%%%%%%%%%%%%%%%%%
\begin{table}[ht]
%\caption{Helical}
\begin{tabular}{ |l|c|c|c| }
      \hline
      & \multicolumn{2}{|c|}{$\bt{h}$ coplanar $\bt{d}(\vec k)$} & $\bt{h} \perp \bt{d}$ \\
      \hline
      $h / \Delta_0$                    & $M^{\text{SC}}$ & $M^{\text{SC}*}$        & $M^{\text{SC}}$                \\
      \hline
      a) $h < \Delta$       & $0.45 M^{\text{N}}$ & $0.46 M^{\text{N}}$               & $M^{\text{N}}$                 \\
      b) $h \approx \Delta$ & $0.56 M^{\text{N}}$ & $0.70 M^{\text{N}}$               & $M^{\text{N}}$                 \\
      c) $h > \Delta$       & $M^{\text{N}}$           & $M^{\text{N}}$                          & $M^{\text{N}}$                 \\
      \hline
\end{tabular}
%\label{table:single-layer-helical}
\caption{Summary of the induced magnetization $M^{\text{SC}}$ for a monolayer with helical superconductivity. Results are for two directions of the magnetic exchange field in three different regimes of amplitude for $\bt{h}$ with respect to the superconducting gap energy $\Delta$, i.e. $h < \Delta$, $ h\approx\Delta$, and $h > \Delta$, respectively.}
\label{table:single-layer-helical}
\end{table}

\subsection{Layer dependent magnetic exchange}

Let us consider a bilayer geometry which is the minimal configuration that allows us to investigate the consequence of a nonuniform magnetic exchange field. Having two distinct layers, one can introduce a layer dependent exchange that can locally mimic the behavior of the induced spin-polarization in a given heterostructure in close proximity to the interface. Hence, we consider both layers to be superconducting, and the exchange magnetic fields have amplitude $h_1$ and $h_2$, respectively. The energy scale of the superconducting order parameters is $\Delta_{1(2)}$. Here, we focus again on the induced magnetic response in each layer by evaluating the spin-polarization for the ground state.

Firstly, we discuss the regime of small exchange amplitude (see rows (a)-(c) in Tables~\ref{table:bi-layer-chiral} and \ref{table:bi-layer-helical}) and uniform $\Delta$.
These regimes are appropriate for a magnetic proximity with the magnetization penetrating inside the superconductor, with the gap keeping about its bulk value.
In the case of uniform $h$ the results substantially reproduce the effects of the single-layer system considered above. In fact, for $\bt {h} \perp \bt{d}$ in both chiral and helical configurations we obtain $M^{\text{SC}} \approx M^{\text{N}}$. Interestingly, when the exchange field is not uniform, instead of following the $\bt{h}(z)$ profile, the magnetization gets smoothed out, such that it has a smaller gradient compared to that in the normal state.
This can be qualitatively ascribed to the fact that Cooper pairs can have a larger interlayer tunneling probability compared to the single electrons in the normal state.  
Another nonstandard behavior is obtained in the regime of strong gradient and large amplitude of the exchange field. There, the magnetization can have an amplitude that is even greater than that of the normal state, as reported in (d)-(f) for $\bt{h} \perp \bt{d}$ in Tables~\ref{table:bi-layer-chiral} and \ref{table:bi-layer-helical}. This implies that a change in the number of electron pairs is in place when $\bt{h}$ is larger than $\Delta$, that allows to have an enhancement of the spin polarization.   
%This behavior is schematically reported in the
%rows (a)-(c) in the Tables~\ref{table:bi-layer-chiral} and ~\ref{table:bi-%layer-helical} for comparison.

Next, we consider the configuration $\bt{h} \parallel \bt{d}$.
The first relevant difference with the previous case is that a variation of the exchange field can lead to a spin polarization with opposite signs in the two layers. This behavior is reminiscent of the oscillation of the magnetization found in the multi-layer system and generally indicates a tendency of the spin-triplet superconductor to exhibit a diamagnetic response when the exchange field is nonuniform.
The helical case, however, never shows a sign change, setting a clear difference between a $\bt{d}$-vector lying in a plane or with a axial configuration. 
Since in the weak gradient regime (row (c) Table~\ref{table:bi-layer-helical}) one can reach a vanishing value of the spin-polarization, it is plausible to expect that a weak sign changing profile of the spin-polarization can be obtained in a multi-layered configuration.

This trend is consistent with the observation of a more pronounced enhancement of $M^{\text{SC}}-M^{\text{N}}$ in the chiral spin-triplet superconductor.

%%%%%%%%%%%%%%%%%%%%%%%%%%%%%%%%%%%%%%%%%
%%%%% Chiral %%%%%%%%%%%%%%%%%%%%%%%%%%%%
%%%%%%%%%%%%%%%%%%%%%%%%%%%%%%%%%%%%%%%%%
\begin{table}[ht]
%\caption{Chiral}
\begin{tabular}{ |r|c|c|c|c| }
      \hline
      \multicolumn{3}{|c|}{ Configurations }
      & $\bt{h} \parallel \bt{d}$ & $\bt{h} \perp \bt{d}$                                                  \\
      \hline
      Case & $ \left(\frac{h_1}{\Delta_0}, \frac{h_2}{\Delta_0}\right) $           & $\left(\frac{\Delta_1}{\Delta_0}, \frac{\Delta_2}{\Delta_0}\right) $ & $\left(\frac{M_1}{M^{\text{N}}}, \frac{M_2}{M^{\text{N}}} \right)$ & $\left(\frac{M_1}{M^{\text{N}}}, \frac{M_2}{M^{\text{N}}} \right)$ \\
      \hline
      a)   & $(0.6, 0.6)$                         & $(1, 1)$                            & $(0, 0)$               & $(1, 1)$               \\
      b)   & $(0.4, 0.6)$                         &   $(1, 1)$                            & $(-0.2, 0.15)$         & $(0.98, 0.99)$         \\
      c)   & $(0.2, 0.6)$                         & $(1, 1)$                            & $(-0.8, 0.3)$          & $(0.80, 0.92)$         \\
      \hline
      d)   & $(0.1, 5)$                           & $(1, 0)$                            & $(0.65, 1)$            & $(1.03, 1)$            \\
      e)   & $(0.1, 20)$                          & $(1, 0)$                            & $(1.5, 1)$             & $(0.96, 1)$            \\
      f)   & $(0.1, 5)$                           & $(1, 0.2)$                          & $(0.53, 0.98)$         & $(1.07, 1)$            \\
      \hline
\end{tabular}
\caption{Summary of the results for the bilayer with chiral superconductivity.}
\label{table:bi-layer-chiral}
\end{table}
  
%\vspace{10pt}
      
%%%%%%%%%%%%%%%%%%%%%%%%%%%%%%%%%%%%%%%%%        
% Helical
\begin{table}[ht]
%\caption{Helical}
\begin{tabular}{ |r|c|c|c|c| }
      \hline
      \multicolumn{3}{|c|}{ Configurations }
      & $\bt{h}$ coplanar $\bt{d}$ & $\bt{h} \perp \bt{d}$                                                 \\
      \hline
      Case & $ \left(\frac{h_1}{\Delta_0}, \frac{h_2}{\Delta_0}\right) $             & $\left(\frac{\Delta_1}{\Delta_0}, \frac{\Delta_2}{\Delta_0}\right) $ & $\left( \frac{M_1}{M^{\text{N}}}, \frac{M_2}{M^{\text{N}}} \right)$& $\left( \frac{M_1}{M^{\text{N}}}, \frac{M_2}{M^{\text{N}}} \right)$ \\
      \hline
      a)   & $(0.6, 0.6)$                           & $(1, 1)$                            & $(0.55, 0.55)$         & $(1, 1)$               \\
      b)   & $(0.4, 0.6)$                           & $(1, 1)$                            & $(0.43, 0.60)$         & $(0.98, 0.99)$         \\
      c)   & $(0.2, 0.6)$                           & $(1, 1)$                            & $(0., 0.63)$           & $(0.80, 0.92)$         \\
      \hline
      d)   & $(0.1, 5)$                             & $(1, 0)$                            & $(0.94, 1)$            & $(1.03, 1)$            \\
      e)   & $(0.1, 20)$                            & $(1, 0)$                            & $(1.19, 1)$            & $(0.96, 1)$            \\
      f)   & $(0.1, 5)$                             & $1, 0.2$                          & $(0.84, 1)$            & $(1.07, 1)$            \\
\hline
\end{tabular}
\caption{Summary of the results for the bilayer with helical superconductivity.}
\label{table:bi-layer-helical}
\end{table}
%%%%%%%%%%%%%%%%%%%%%%%%%%%%%%%%%%%%%%%%%%%

%%%%%%%%%%%%%%%%%%%%%%%%%
%%%%%%%%% AB INITIO    %%
%%%%%%%%%%%%%%%%%%%%%%%%%

\section{Ab-initio analysis}

In this Section we provide the details of the ab-initio computation with the evolution of the magnetic instability in Sr$_2$RuO$_4$ as a function of Coulomb interaction and strain.

\subsection{Computational details}

We perform spin polarized first-principles density functional calculations within the LSDA (Local Spin Density Approximation) \cite{Kohn64,Kohn64a} by using the plane wave VASP \cite{VASP} DFT package and the Perdew-Zunger \cite{Perdew} parametrization of the Ceperly-Alder data \cite{Ceperley} for the exchange-correlation functional. 
The choice of the LSDA exchange functional is suggested by a recent paper \cite{Etz12} where the Generalized Gradient Approximation (GGA) is shown to be an approximation less accurate than LSDA for SrRuO$_3$. 
The core and the valence electrons were treated within the projector augmented wave (PAW) method \cite{BlochlPAW} and a cutoff of 430~eV for the plane wave basis. An 8$\times$8$\times$1 $k$-point grid is used for the full relaxation of the bulk orthorhombic phase with spin orbit coupling.  We use
a 10$\times$10$\times$1 $k$-point grid centered at the Gamma point for
the determination of the total energy of Sr$_2$RuO$_4$ and the full relaxation of the heterostructure. In all cases the tetrahedron method with Bl\"{o}chl corrections \cite{BlochlCORR} was used for the Brillouin zone integrations. We optimize
the internal degrees of freedom by minimizing the
total energy to be less than 10$^{-5}$~eV and the remaining
forces to be less than 5~meV/{\AA}.

To catch the magnetic behavior, the Hubbard $U$ effects at the Ru sites for 4$d$ orbitals was included in the LSDA+$U$ \cite{Anisimov91,Anisimov93} approach
using the rotationally invariant scheme proposed in Ref. \onlinecite{Anisimov95} and implemented in VASP \cite{VASP}.
We denote by $U_{113}$ and $J_{H,{113}}$ the Coulomb repulsion and the Hund parameter of {\sro}, while $U_{214}$ and $J_{H,{214}}$ for the {\sroa}.
In this study we fixed $U_{113}=1$eV for the Ru 4$d$ states of {\sro}, while we use $J_{H,{113}}$=0.15 $U_{113}$.
We keep $J_{H,{214}}$=0 and we tune from 0 to 1.5~eV the Coulomb repulsion in {\sroa}.
These values are common to all the supercell calculations presented in this work.

The study of SrRuO$_3$ under strain is made with the in-plane lattice constant of the {\sroa}.
A small values of $U$ was used in the bulk to reproduce the correct magnetization in DFT, but in extreme strain conditions one does not have a reference of $U$ to describe the corresponding physical condition.

In order to simulate thin films of SrRuO$_3$ grown on Sr$_2$RuO$_4$ substrate, we fix the in-plane lattice parameter $a$ with the experimental value of the Sr$_2$RuO$_4$ bulk and the atomic positions of the {\sroa} phase are fixed to the bulk values.
We use the value of $c$ such that the volume of SrRuO$_3$ is equal to the bulk volume because the theoretical volume underestimates the experimental one.
We discuss the structural and electronic properties of heterostructures made of {\sroa} and {\sro}. We will investigate the supercell composed by three {\sroa} and five {\sro} layers. The focus is on the superlattice structures with three inequivalent {\sroa} layers and three inequivalent layers of SrRuO$_3$.
Being the magnetic order in SrRuO$_3$ dominating, we first study the magnetic order in the SrRuO$_3$ side of the heterostructure fixing $U_{214}$=0.
After that, we add the Coulomb repulsion in {\sroa} to investigate the magnetic instability of the Sr$_{2}$RuO$_{4}$ side of the supercell.
Structural optimization was performed in LSDA+$U$ approximation
separately for the different magnetic cases.

\subsection{Magnetic instability in {\sroa}}

We discuss the role of the Coulomb interaction and the strain in {\sroa} system together with the possible magnetic instabilities and their character.

The Sr$_{2}$RuO$_{4}$ compound is paramagnetic in the bulk. We compare the energy of the ferromagnetic (FM), paramagnetic (PM) and antiferromagnetic (AFM) phase in terms of $U^{214}$ from 0 to 1~eV and by varying the in-plane lattice constant from -3\% to +3\% . 
The outcome is reported in Fig. \ref{fig:figureMagnPhase}.
%The internal degrees of freedom are not fixed.
The AFM phase is G-type, as it can be also found in the ultrathin SrRuO$_3$.
The system is paramagnetic for small values of $U$ and it is close to both AFM and FM magnetic phase for compressive and tensile strain, respectively.

The rotations of the octahedra are not stabilized by the strain. Because there are no rotations, the most important effect is the dependence of the
hopping parameters on the Ru-Ru distance.
When the lattice constant gets larger, the hopping decreases and the role of electron correlations becomes more relevant.
This explains the magnetic instability at larger in-plane lattice constant ($a$) as a function of $U$.
If we impose the rotations of the octahedra and the consequent reduction
of the bandwidth, there is a critical value of the rotation beyond which Sr$_{2}$RuO$_{4}$ becomes ferromagnetic at any small value of the Coulomb interaction.
The magnetic properties in Fig. \ref{fig:figureMagnPhase} do not change qualitatively if we apply spin-orbit or by assuming that the atomic positions are not relaxed.

Now, we fix the atomic positions to the experimental values and the volume
to study more accurately the transition from the PM to the FM phase.

\begin{figure}[!ht]
\centering
\includegraphics[width=0.5\textwidth]{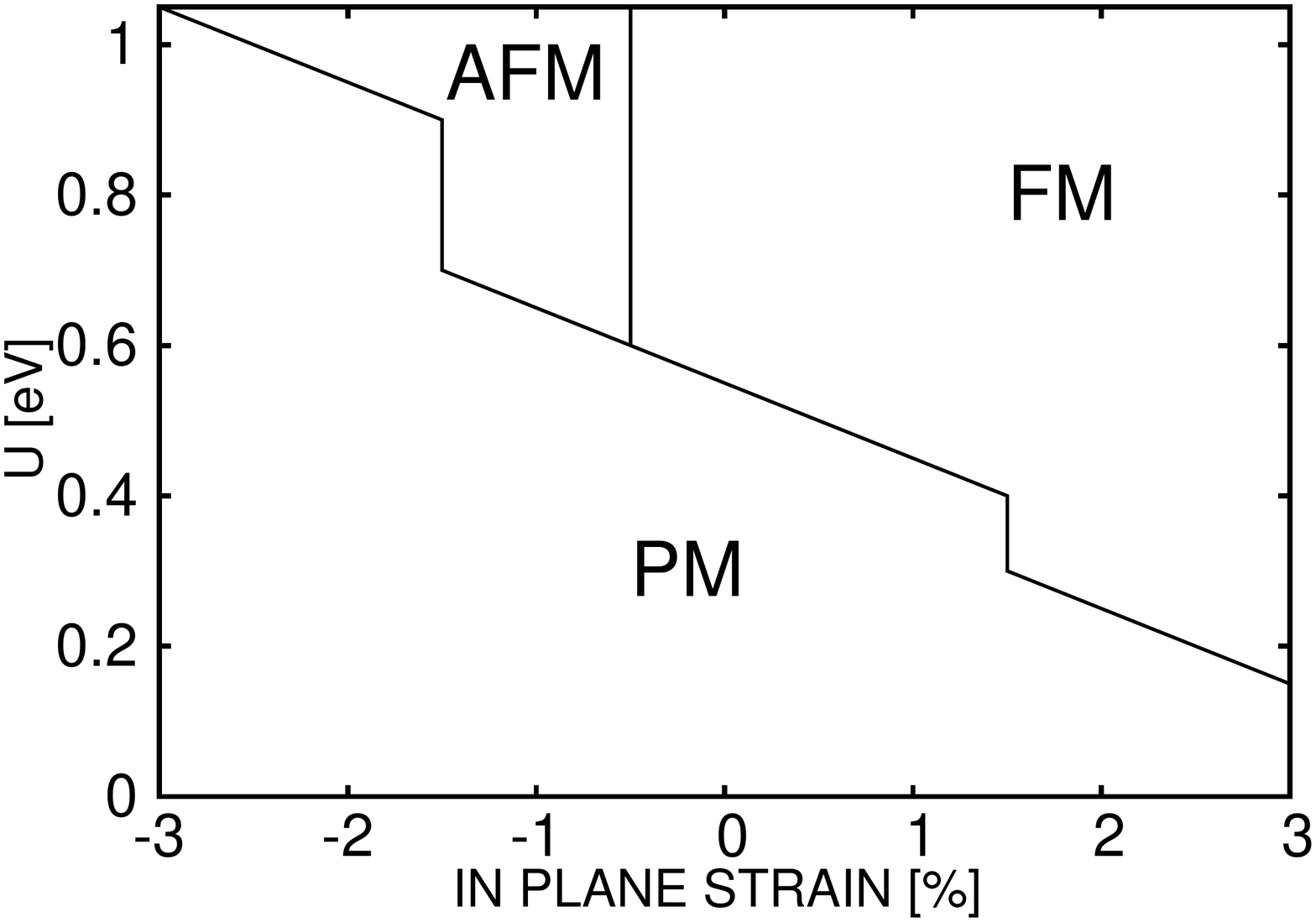}
\caption{Magnetic phase diagram of bulk Sr$_2$RuO$_4$ as a function of the Coulomb repulsion and in-plane strain.}
\label{fig:figureMagnPhase}
\end{figure}
\begin{figure}[!ht]
\centering
\includegraphics[width=0.4\textwidth]{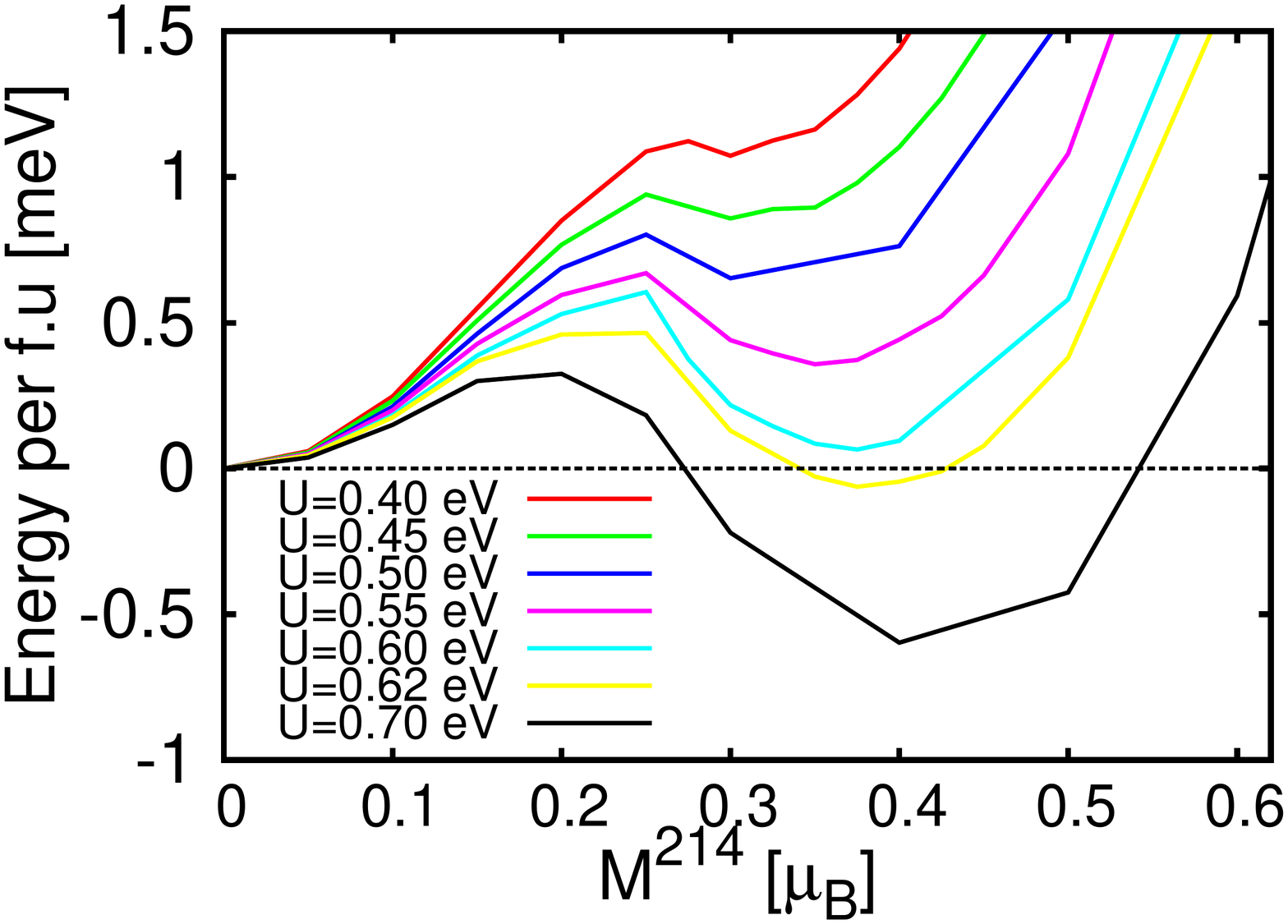}
\includegraphics[width=0.4\textwidth]{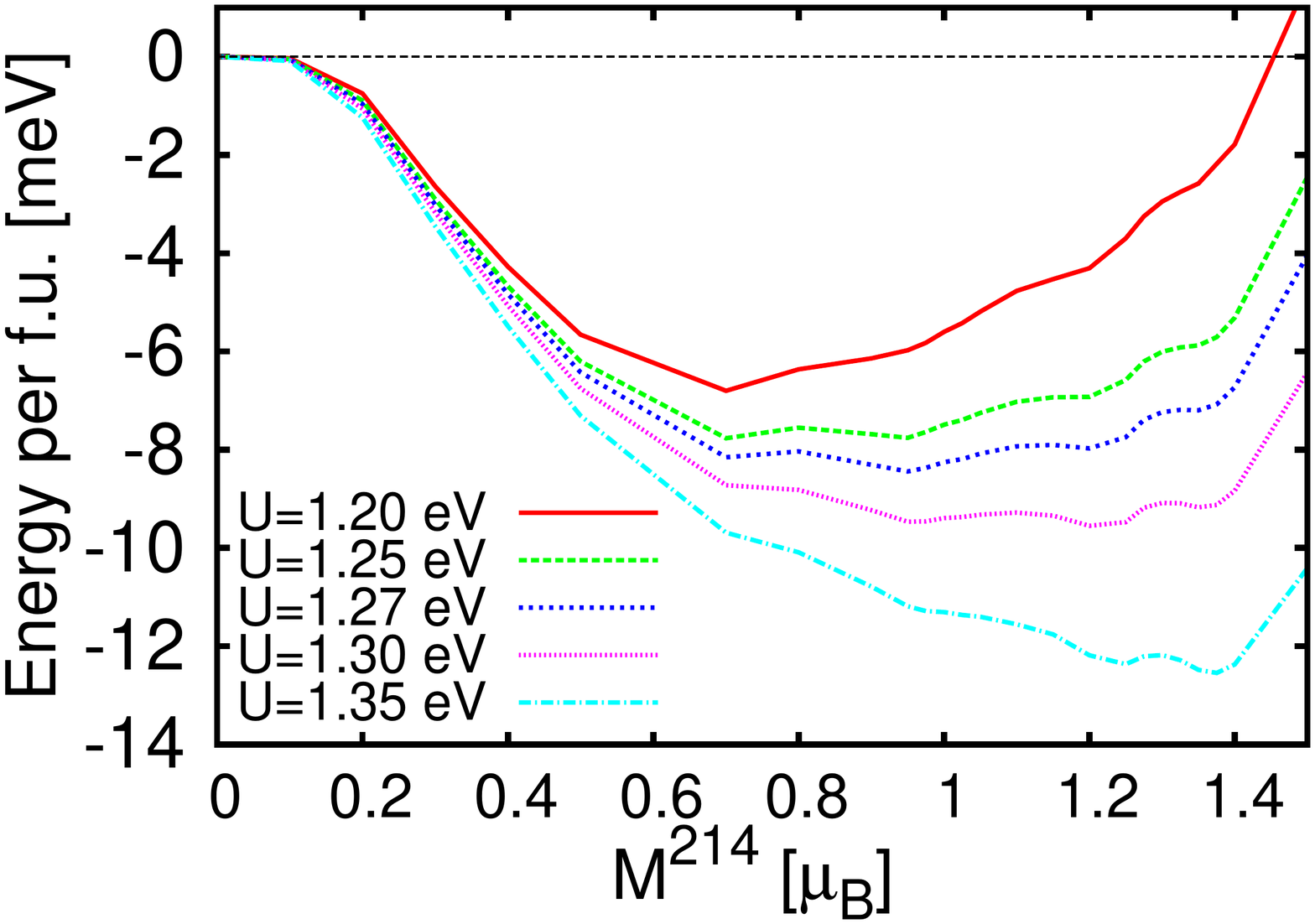}
\caption{(Color online) Energy as function of the magnetic moment of the bulk Sr$_2$RuO$_4$
for different values of the Coulomb repulsion $U$. In the top panel we show the
presence of the ferromagnetic minimum close to the paramagnetic minimum thanks to the VHS in the d$_{xy}$ band.
In the bottom panel we can observe a first order transition between
two ferromagnetic minima thanks to the VHS in the d$_{{\gamma}z}$ bands. The energy of the paramagnetic phase
is the reference point.
}
\label{fig:M214_EM_gammaz}
\end{figure}

We calculate in detail the total energy as a function of the magnetization in Fig. \ref{fig:M214_EM_gammaz}
for the values of $U$ close to the magnetic transition.
In the top panel we can observe the double minima curve giving rise to a first order magnetic transition that is typical in a metamagnetic system.
The spin-polarized state becomes lower in energy when increase $U$ and becomes the ground state at U=0.61~eV.

%%%%%%%%%%%%%%%%%%%%%%%%%%%%%%%%%%%%%%%
%%%%%%%%%%%%%%%%%%%%%%%%%%%%%%%%%%%%%%%
%%%%%%%%%%%%%%%%%%%%%%%%%%%%%%%%%%%%%%%

\end{document}